\documentstyle[aps,12pt,preprint]{revtex}

\tightenlines 

\begin{document}

\title{A Note on the Lorentz Transformations for Photon}

\author{Valeri V. Dvoeglazov and J. L. Quintanar Gonz\'alez}

\maketitle

\begin{center}{\it Universidad de Zacatecas\\ Apartado Postal 636, Suc. UAZ, C. P. 98062, Zacatecas, Zac., M\'exico\\ e-mail: valeri@cantera.reduaz.mx, \, el\_leo\_xyz@yahoo.com.mx }
\end{center}

\begin{abstract}
\noindent 
We discuss transormation laws of electric and magnetic fields under Lorentz transformations, deduced from the Classical  Field Theory. It is found that we can connect the resulting expression for a bivector formed with those fields, with the expression deduced from the Wigner transformation rules for spin-1 functions of {\it massive} particles. This mass parameter 
should be interpreted because the constancy of speed of light forbids the existence of the photon mass.
\end{abstract}

\section{Introduction}

Within the Classical Electrodynamics (CED) we can obtain transformation rules for electric and magnetic fields when we pass from one frame to another which is moving with respect to the former with constant velocity; in other words, we can obtain the relationships between the fields under Lorentz transformations or {\it boosts}. On the other hand we have that electromagnetic waves are constituted of ``quanta" of the fields, which are called {\it photons}. It is usually accepted that photons do not have mass. Furthermore, the photons are the particles which can be in the eigenstates of helicities $\pm 1$. The dynamics of such fields is described by the Maxwell equations on the classical level. On the other hand, we know the {\it Weinberg-Tucker-Hammer formalism}~\cite{wein,dvoe} which describe spin-1 massive particles. The massless limit of the Weinberg-Tucker-Hammer formalism can be well-defined in the light-cone basis~\cite{saw}. 

In this work we show how these two viewpoints can be related.

\section{The Lorentz Transformations for Electromagnetic Field Presented by Bivector}

In  the Classical Electrodynamics we know the following equations to transform electric and magnetic fields under Lorentz transformations~\cite{landau}
\begin{eqnarray}
{\bf E'}&=&\gamma({\bf E}+{{\bbox \beta} \times B})-\frac{\gamma^2}{\gamma+1}{\bbox \beta}({{\bbox \beta} \cdot E}) \label{campoE}\\
{\bf B'}&=&\gamma({\bf B}-{{\bbox \beta} \times E})-\frac{\gamma^2}{\gamma+1}{\bbox \beta}({{\bbox \beta} \cdot B})\label{campoB}
\end{eqnarray}
where $\gamma=1/\sqrt{1-\frac{v^2}{c^2}}$, ${\bbox\beta}= {\bf v}/c$; ${\bf E}, {\bf B}$ are the field in the originary frame of reference, ${\bf E}^\prime$, ${\bf B}^\prime$ are the fields in the transformed frame of reference. In the Cartesian component form we have
\begin{eqnarray}
E^{i^\prime}&=&\gamma(E^i+\epsilon^{ijk}\beta^j B^k)-\frac{\gamma^2}{\gamma+1} \beta^i \beta^j E^j\label{campoE2}\\
B^{i^\prime}&=&\gamma(B^i-\epsilon^{ijk}\beta^j E^k)-\frac{\gamma^2}{\gamma+1} \beta^i \beta^j B^j
\label{campoB2} \end{eqnarray}
Now we introduce a particular representation of ${\bf S}$ matrices (generators of rotations for spin 1): $({\bf S}^i)^{jk}=-i\epsilon^{ijk}$, i.e.
\begin{equation}
S_x=\left(\begin{array}{ccc}0&0&0\\0&0&-i\\0&i&0\end{array}\right),\hspace{2mm}
S_y=\left(\begin{array}{ccc}0&0&i\\0&0&0\\-i&0&0\end{array}\right),\hspace{2mm}
S_z=\left(\begin{array}{ccc}0&-i&0\\i&0&0\\0&0&0\end{array}\right).
\end{equation}
Using the relation $\epsilon^{ijk}\epsilon^{lmk}=\delta^{il}\delta^{jm}-\delta^{im}\delta^{jl}$ (the Einstein sum rule on the repeated indices is assumed),  we have for an arbitrary vector ${\bf a}$:
\begin{equation}
({\bf S\cdot a})^2_{ij}={\bf a}^2\delta^{ij}-a^i a^j\,.
\label{Spuntoa} \end{equation}
So with the help of the ${\bf S}$ matrices we can write (\ref{campoE2},\ref{campoB2}) like
\begin{eqnarray}
E^{i^\prime}=\gamma(E^i-i(S^j)^{ik}\beta^j B^k)-\frac{\gamma^2}{\gamma+1}[{\bbox\beta}^2 \delta^{ij}-({\bf S\cdot {\bbox\beta})}^2_{ij}] E^j\label{EconS}\,,\\
B^{i^\prime}=\gamma(B^i+i(S^j)^{ik}\beta^j E^k)-\frac{\gamma^2}{\gamma+1}[{\bbox\beta}^2 \delta^{ij}-({\bf S\cdot {\bbox\beta})}^2_{ij}] B^j\,,
\label{BconS} \end{eqnarray}
or
\begin{eqnarray}
{\bf E'}=\{\gamma-\frac{\gamma^2}{\gamma+1}[{\bbox\beta}^2-({\bf S\cdot {\bbox\beta})}^2]\}{\bf E}-i\gamma({\bf S\cdot {\bbox\beta}}){\bf B}\label{EconS2}\\
{\bf B'}=\{\gamma-\frac{\gamma^2}{\gamma+1}[{\bbox\beta}^2-({\bf S\cdot {\bbox\beta})}^2]\}{\bf B}+ i\gamma({\bf S\cdot {\bbox\beta}}){\bf E}\,.
\label{BconS2} \end{eqnarray}
In the matrix form we have:
\begin{equation}
\left(\begin{array}{c} {\bf E'}\\{\bf B'}\end{array}\right)=
\left(\begin{array}{cc} \gamma-\frac{\gamma^2}{\gamma+1}[{\bbox\beta}^2-({\bf S\cdot {\bbox\beta})}^2]&-i\gamma({\bf S\cdot {\bbox\beta}})\\
i\gamma({\bf S\cdot {\bbox\beta}})&\gamma-\frac{\gamma^2}{\gamma+1}[{\bbox\beta}^2-({\bf S\cdot {\bbox\beta})}^2]\end{array}\right)
\left(\begin{array}{c} {\bf E}\\{\bf B}\end{array}\right).
\label{matrizEB} \end{equation}
Now we introduce the unitary matrix $U=\frac{1}{\sqrt{2}}\left(\begin{array}{rr} 1&i\\1&-i\end{array}\right)$ which satisfies $U^{\dagger}U=1$. Multiplying the equation (\ref{matrizEB}) by this matrix we have
\begin{equation}
U\left(\begin{array}{c} {\bf E'}\\{\bf B'}\end{array}\right)=
U\left(\begin{array}{cc} \gamma-\frac{\gamma^2}{\gamma+1}[{\bbox\beta}^2-({\bf S\cdot {\bbox\beta})}^2]&-i\gamma({\bf S\cdot {\bbox\beta}})\\
i\gamma({\bf S\cdot {\bbox\beta}})&\gamma-\frac{\gamma^2}{\gamma+1}[{\bbox\beta}^2-({\bf S\cdot {\bbox\beta})}^2]\end{array}\right)U^{\dagger}U
\left(\begin{array}{c} {\bf E}\\{\bf B}\end{array}\right),
\label{matrizEB2} \end{equation}
which can be reduced to
\begin{equation}
\left(\begin{array}{c} {\bf E'}+i{\bf B'}\\{\bf E'}-i{\bf B'}\end{array}\right)=
\left(\begin{array}{cc} 1-\gamma({\bf S\cdot {\bbox\beta}})+\frac{\gamma^2}{\gamma+1}({\bf S\cdot {\bbox\beta})}^2&0\\
0&1+\gamma({\bf S\cdot {\bbox\beta}})+\frac{\gamma^2}{\gamma+1}({\bf S\cdot {\bbox\beta})}^2 \end{array}\right)
\left(\begin{array}{c} {\bf E}+i{\bf B}\\{\bf E}-i{\bf B}\end{array}\right).
\label{matrizEB3} \end{equation}
Now, let us take into account that ${\bbox\beta}$-parameter is related to the  momentum and the energy in the following way: when we differentiate
$E^2-{\bf p}^2 c^2=m^2 c^4$ we obtain $2E\mbox{d}E-2c^2{\bf p}\cdot\mbox{d}{\bf p}=0$, hence $\frac{\mbox{d}E}{\mbox{d}{\bf p}}=c^2 \frac{{\bf p}}{E}={\bf v}= c{\bbox \beta}$. Then, we set $\gamma =\frac{E}{mc^2}$, where we must interpretate $m$ like some mass parameter (as in~\cite[p.43]{ryder}). It is rather related not to the photon mass  but to the particle mass, with which we associate the second frame (the energy and the momentum as well).
So, we have
\begin{equation}
\left(\begin{array}{c} {\bf E'}+i{\bf B'}\\{\bf E'}-i{\bf B'}\end{array}\right)=
\left(\begin{array}{cc} 1-\frac{({\bf S\cdot p})}{mc}+\frac{({\bf S\cdot p)}^2}{m(E+mc^2)}&0\\
0&1+\frac{({\bf S\cdot p})}{mc}+\frac{({\bf S\cdot p)}^2}{m(E+mc^2)} \end{array}\right)
\left(\begin{array}{c} {\bf E}+i{\bf B}\\{\bf E}-i{\bf B}\end{array}\right).
\label{ecbivector} \end{equation}
Note that we have started from the transformation equations for the fields, which do not involve any mass and, according to the general wisdom, they should describe massless particles. 

\section{The Lorentz transformations for Massive Spin-1 Particles in the Weinberg-Tucker-Hammer Formalism}

When we want to consider Lorentz transformations  and derive relativistic quantum equations for quantum-mechanical state functions, we first have to work with the  representations of the quantum-mechanical Lorenz group. These representations have been studied by E. Wigner~\cite{wigner}. In order to consider the theories with definite-parity solutions of the corresponding dynamical equations  (the 'definite-parity' means that the solutions are the eigenstates of the spatial reflection operation), we have to look for a function formed by two components (called the ``right" and ``left" components),ref~\cite{ryder}. According to the Wigner rules, we have the following expressions
\begin{eqnarray}
\phi_R(p^{\mu})=\Lambda_R({p}^{\mu}\leftarrow\stackrel{0}{p^{\mu}})\phi_R(\stackrel{0}{p^{\mu}})\label{phiR}, \\
\phi_L(p^{\mu})=\Lambda_L({p}^{\mu}\leftarrow\stackrel{0}{p^{\mu}})\phi_L(\stackrel{0}{p^{\mu}})\,\label{phiL}
\end{eqnarray}
where $\stackrel{0}{p^{\mu}} =(E, {\bf 0})$ is the 4-momentum at rest, $p^\mu$ is the 4-momentum in the 2nd frame (where a particle has 3-momentum ${\bf p}$, $c=\hbar =1$).
In the case of spin $S$, $\psi=\pmatrix{\phi_R(p^{\mu})\cr \phi_L(p^{\mu})}$ is called the Weinberg $2(2S+1)$ function~\cite{wein}. Let us consider the case of  $S=1$. The matrices  $\Lambda_{R,L}$ are then the matrices of the $(1,0)$, $(0,1)$ representations of the Lorentz group, respectively. Their explicite forms are (${\bbox \phi} = {\bf n} \phi$)
\begin{equation}
\Lambda_{R,L}=\exp(\pm \bf S\cdot {\bbox\phi})=1+({\bf S\cdot\hat{n}})^2\left[\frac{\phi^2}{2!}+\frac{\phi^4}{4!}+\frac{\phi^6}{6!}+...\right]\\
\pm{\bf S\cdot\hat{n}}\left[\frac{\phi}{1!} +\frac{\phi^3}{3!}+\frac{\phi^5}{5!}+...\right]\,,
\label{lambdas}
\end{equation}
or
\begin{equation}
\exp(\pm{\bf S\cdot \phi})=1+({\bf S\cdot\hat{n}})^2(\cosh\phi -1)
\pm({\bf S\cdot\hat{n}})\sinh\phi\,.
\end{equation}

If we introduce the parametrizations $\cosh\phi =\frac{E}{m},\hspace{2mm}\sinh\phi =\frac{\mid{\bf p}\mid}{m},\hspace{2mm}{\bf \hat{n}}=\frac{{\bf p}}{\mid{\bf p}\mid}$, see~\cite[p.39-43]{ryder}, $c=\hbar=1$, we obtain
\begin{eqnarray}
\Lambda_R ({p}^{\mu}\leftarrow\stackrel{0}{p^{\mu}}) &=& 1+\frac{{\bf S\cdot p}}{m}+\frac{({\bf S\cdot p})^2}{m(E+m)},\\
\Lambda_L ({p}^{\mu}\leftarrow\stackrel{0}{p^{\mu}}) &=& 1-\frac{{\bf S\cdot p}}{m}+\frac{({\bf S\cdot p})^2}{m(E+m)}\,.
\label{lambdas2}
\end{eqnarray}
Thus, the equations (\ref{phiR}, \ref{phiL}) are written as
\begin{eqnarray}
\phi_R(p^{\mu})=\left\{1+\frac{{\bf S\cdot p}}{m}+\frac{({\bf S\cdot p})^2}{m(E+m)}\right\}\phi_R(\stackrel{0}{p^{\mu}})\label{phiR1}, \\
\phi_L(p^{\mu})=\left\{1-\frac{{\bf S\cdot p}}{m}+\frac{({\bf S\cdot p})^2}{m(E+m)}\right\}\phi_L(\stackrel{0}{p^{\mu}}).\label{phiL1}
\end{eqnarray}
If we compare the equations (\ref{phiR1},\ref{phiL1}) with the equation (\ref{ecbivector}) we see that ${\bf E}-i{\bf B}$ can be considered as $\phi_R$, ${\bf E}+i{\bf B}$ can be considered as $\phi_L$.
 
\section{Conclusions}

We have found that when we introduce a mass parameter in the equation (\ref{matrizEB3}) we can make  the equation (\ref{ecbivector}) and the equations (\ref{phiR}, \ref{phiL}) to coincide. This result suggests we have to attribute the mass parameter to the frame and not to the electromagnetic-like fields. This should be done in order to preserve the postulate which states that all inertial observers must measure the same speed of light. Moreover, our considertion illustrates a sutuation in which we have to distinguish between passive and active transformations.

We are grateful to Sr. Alfredo Casta\~neda for discussions in the classes of quantum mechanics at the University of Zacatecas.

\end{document}